\documentclass[11pt]{article}

\newcommand{\be}{\begin{equation}}
\newcommand{\ee}{\end{equation}}
\newcommand{\ba}{\begin{eqnarray}}
\newcommand{\ea}{\end{eqnarray}}
\newcommand{\nn}{\nonumber}

\newcommand{\MSsch}{{\overline{\rm MS}}}
\newcommand{\ice}[1]{\relax}
\newcommand{\als}{{\bar{\alpha}}_s}
\begin{document}
\begin{center}
\vspace*{0.5in}
{\Large \bf Renormalization group analysis\\
of the $\tau$-lepton decay within QCD\footnote{Electronic 
version of the published paper,
the journal reference: Yad.~Fiz. 54, 1114 (1991) and \\
Z.~Phys. C53, 461 (1992).}
}
\vspace{0.3in}

{\large \bf A.A.~Pivovarov}\\
\vspace{0.1in} 
{\it Institute for
Nuclear Research of the Russian Academy of Sciences \\ 
Moscow 117312, Russia}
\vspace{0.5in} 

{\bf Abstract} \end{center}
\vspace{-0.1in} 
\noindent A technique to sum up the regular corrections appearing 
under the analytic continuation from the spacelike momentum region to the
timelike one is proposed. A perturbative part of the inclusive
semileptonic decay width of the $\tau$-lepton in analyzed in detail.

\vskip 1.5cm 

At present the theory of electroweak interactions is tested with
truly great accuracy due to creation of the modern
experimental machines~\cite{1} and to the happy possibility to fulfill the
reliable theoretical calculations within the perturbative framework in
the gauge coupling constants which are small~\cite{2}. The standard model
as a whole however can not be checked with the same impressive
accuracy because of uncertainties introduced by the strong
interactions into the theoretical predictions at low energies.

Inclusive process $\tau\rightarrow\nu+hadrons$ is one of the most
convenient channels to explore strong interactions in the low
energy region and, in particular, to determine the numerical value of
the strong coupling constant $\als$. The reason consists in two
crucial observations: first, the problems of describing the final
hadronic states disappear while calculating the full width of the
decay and, second, the nonperturbative corrections to the decay rate
are small due to some specific features of the process 
kinematics~\cite{3,4}.

In the present note we estimate the accuracy of the calculation for
the perturbative part of the  full decay
width $\Gamma(\tau\rightarrow\nu+hadrons)$ when the analytic
continuation effects are summed up in all orders of the perturbative
$\als$-expansion.

An expression for the full width of the decay
$\tau\rightarrow\nu+hadrons$ can be written down in the 
form~\cite{5} 
\be
\label{1}
\Gamma(\tau^{-})=
\Gamma(\tau^{-}\rightarrow\nu_{\tau}e^{-}\bar{{\nu}_e})(4.9728+12r_H)
\ee
where the perturbative contribution of the strong interactions
$r_H$ which we are interested in is given by the following spectral
integral 
\be
\label{2}
r_H=2\int_0^{m_\tau^2}r(s) \left(1 - {s\over m_\tau^2}\right)^2
\left(1+2{s\over m_\tau^2}\right){ds\over m_\tau^2}\, .
\ee 
The spectral density $r(s)$ is determined by the correlator of weak hadronic
currents (see refs.~\cite{3,4,5} for more details) and is connected with the
theoretically calculable quantity $d(Q^2)$~\cite{6} by 
the K\"allen-Lehmann representation
\be
\label{3}
d(Q^2)=Q^2\int{r(s)ds\over{(s+Q^2)^2}}\, .
\ee
The perturbative expansion of the function $d(Q^2)$ in the running
strong interactions coupling constant $h(Q^2)=\als(Q^2)/4\pi$ looks like
\be
\label{4}
d(Q^2) = h(Q^2)(1+a_1 h(Q^2)+a_2 h(Q^2)^2 +a_3 h(Q^2)^3+\ldots )
\ee
where the coefficients $a_1,~a_2,~a_3$ are determined by the
divergent parts of the three, four, five loop Feynman diagrams
correspondingly.

To calculate the full decay width through eqs.~(\ref{1}) and~(\ref{2}) it
is necessary to find the moments of the spectral density $r(s)$
\[
r_n={m_\tau}^{-2(n+1)}\int_0^{m_\tau^2}r(s){s^n}ds\, .
\]

An accurate definition of such moments can be given with the help
of the integration contour which goes
along the positive semi axis coinciding with the cut
in the complex $Q^2$ plane and closes itself far enough from
all physical singularities~\cite{7,8,9,10}. 
The result for these moments does
not depend on the explicit way of the closing of the contour if it
lies in the analyticity region of the function $d(Q^2)$. Choosing the
circle of the radius $m_\tau^2$ for the closing of the contour we get
the representation for the moments in the following form~\cite{11}
\be
\label{5}
r_n={m_\tau}^{-2(n+1)}\int_0^{m_\tau^2}r(s){s^n}ds=
{1\over 2\pi}\int_{-\pi}^{\pi}p(m_\tau^2e^{i\phi})e^{i(n+1)\phi}d\phi
\ee 
where the function $p(Q^2)$ is given by the expression
\be
\label{6}
p(Q^2)=p(\mu^2)-\int^{Q^2}_{\mu^2}{d\xi\over\xi}d(\xi)\, .
\ee

The running strong coupling constant $h(Q^2)$ satisfies the
standard renormalization group equation
\be
\label{7}
Q^2{d\over dQ^2}h(Q^2)=\beta(h(Q^2))=
-\beta_0 h(Q^2 )^2-\beta_1 h(Q^2)^3-\beta_2 h(Q^2)^4
\ee
and the coefficients of the expansion for the $\beta$-function  are
equal to $\beta_0=9$, $\beta_1=64$, $\beta_2=3863/6$ in the
$\MSsch$-renormalization scheme for $N_F=3$~\cite{13}.

At the contour $Q^2=m_\tau^2e^{i\phi}$, $-\pi<\phi<\pi$ the running
strong coupling constant $h(Q^2)$ becomes the complex function of the
angle $\phi$ and the renormalization group equation takes the form
\be
\label{8}
{\partial h(m_\tau^2e^{i\phi})\over\partial \phi}=
i\beta (h(m_\tau^2e^{i\phi})).
\ee

Setting $h(Q^2)=h(m_\tau^2)(u+iv)^{-1}$ at the contour one gets for
the functions $u(\phi)$ and $v(\phi)$ the system of equations in the
given approximation (\ref{7}) for the $\beta$-function
\ba
\label{9}
{\partial u\over\partial \phi}&=& \beta_1 h^2{v\over{u^2+v^2}}
+ \beta_2 h^3{2uv\over{(u^2+v^2)}^2} \nn \\
&&\\
{\partial v\over\partial \phi}
&=&\beta_0 h+\beta_1 h^2{u\over{u^2+v^2}}
+ \beta_2 h^3{u^2-v^2\over{(u^2+v^2)}^2} \nn
\ea
with the initial
conditions $u(0)=1$, $v(0)=0$.

Formulas (\ref{5},\ref{6},\ref{9}) allow to sum up all 
the regular corrections in $h(m_\tau^2)$ to the moments
$r_n$ of the spectral density $r(s)$ with the renormalization group
technique. The naive expression for the
moments $r_n$ can be obtained by expanding the exact formulas in the
series in $h(m_\tau^2)$ and can be also found by the following simple
trick. The renormalization group properties of the
quantity $r(s)$ up to the order ${\cal O}(h(\mu^2)^3)$ can be explicitly
expressed in the form
\be
\label{10}
r(s)=h(\mu^2)+h(\mu^2)^2\left(a_1+\beta_0 L\right)
+h(\mu^2)^3\left(a_2+(\beta_1+2\beta_0a_1)L
+\beta_0^2\left(L^2-{\pi^2\over 3}\right)\right)
\ee 
where $L=\ln(\mu^2/s)$
and we have taken into account the representation (\ref{3}) and
expression (\ref{4}). After integrating eq.~(\ref{10}) over the interval
$(0,m_\tau^2)$ (more precisely the corresponding function $p(Q^2)$
along the contour in the complex plane of the variable $Q^2$) we get
formally the following expression for the zero moment, for example, at
$\mu^2=m_\tau^2$
\[
r_0=h+h^2(a_1+\beta_0)+
h^3\left(a_2+\left(\beta_1+2\beta_0a_1\right)
+\beta_0^2\left(2-{\pi^2\over 3}\right)\right),
\] 
where we use the simple notation $h=h(\mu^2)$.

It is clear that such an integration generates a
good deal of regular terms connected with the coefficients of the
$\beta$-function which can and must be summed up with the
renormalization group technique~\cite{12}.

A following remark is in order here.  Within the naive approach, 
eq.~(\ref{10}), it is not clear how many terms of the expansion 
in $h(\mu^2)$ must be kept. 
In ref.~\cite{5}, for example, all terms of the same order
as in the expansion for the $d(Q^2)$-function were kept in the
$h$-expansion for the spectral density $r(s)$
in the $\MSsch$-renormalization scheme.  This prescription does
not work if the so called effective renormalization scheme is used for
the function $d(Q^2)$ in which the corrections are absent at all due
to definition ($a_1=0$, $a_2=0$, $a_i=0$ for $i>2$)~\cite{14}. Within our
approach all regular terms determined by the expansion of the
$\beta$-function are summed up and the action instruction is
fully independent of the choice of the renormalization scheme  for the
quantity $d(Q^2)$. After integration by parts in eq.~(\ref{5}) we get the
representation for the moments
\[
r_n={1\over{n+1}}(r_f+\Delta_n),
\] 
where 
\[
r_f={p(-m_\tau^2+i0) -p(-m_\tau^2-i0)\over2\pi i}
\]
\be
\label{11}
={1\over{\beta_0\pi}}\left(\arctan(v/u)+k_1h{v\over{u^2+v^2}}
+k_2h^2{uv\over{(u^2+v^2)}^2}\right),
\ee 
and $k_1=a_1-c_1$,
$k_2=a_2-c_2+c_1^2-a_1c_1$, $c_1=\beta_1/\beta_0$,
$c_2=\beta_2/\beta_0$, the variables $u$ and $v$ in formula~(\ref{11})
denote the values of the solution to the system of equations~(\ref{9}) at
the point $\phi=\pi$, while the quantities $\Delta_n$ are determined
by the integrals
\be
\label{12}
\Delta_n={1\over 2\pi}
\int_{-\pi}^{\pi}d(m_\tau^2e^{i\phi})e^{i(n+1)\phi}d\phi.
\ee

Substituting the solution of the system of equations~(\ref{9}) into 
formulas~(\ref{11}-\ref{12}) one gets after performing the integration
the numerical
values for the moments $r_n$ of the spectral  density $r(s)$ with the
full summation of the analytic continuation effects by the
renormalization group method. The experimental value obtained for the
quantity $r_H$ from the relation~(1) after averaging over several
possible ways of extracting the numerical value for
$\Gamma(\tau\rightarrow\nu+hadrons)$ and $\Gamma(\tau^{-})=
\Gamma(\tau^{-}\rightarrow\nu_{\tau}e^{-}\bar{{\nu}_e})$ is
equal to~\cite{5}
\be
\label{13}
12r_H=0.50\pm 0.07.
\ee
The theoretical
prediction is based on formula~(\ref{2}) and is expressed through the
moments of the spectral density $r(s)$ as follows
\be
\label{14}
r_H=2(r_0-3r_2 +2r_3).
\ee

The naive expression for the quantity $r_H$ up to the order $O(h^3)$
has the form
\be
\label{15}
r_H^N =h_N +h_N^2 \left(a_1 +{19\over12}\beta_0\right)
+h_N^3\left(a_2+\left({19\over12}\beta_1 +2\beta_0 a_1\right)
+\beta_0^2 \left({265\over72}-{\pi^2\over3}\right)\right),
\ee
where $h_N$  denotes the running coupling constant $h$ extracted from
the naive formula~(\ref{15})  and the coefficients $a_1$, $a_2$ are
$a_1=6.564$~\cite{15},  $a_2=102.0$~\cite{16} in the
$\MSsch$-scheme of renormalization.

The results of determination of the running coupling constant $h$
and the QCD-parameter $\Lambda_{\MSsch}$ in the different
approximations in the coupling constant with the help of 
formulas~(\ref{14}) and~(\ref{15}) are given in Tables 1-3.  
The mass scale of QCD $\Lambda_{\MSsch}$ is determined through 
the numerical value of
the running coupling constant $h$ according to the relation
\[
\ln(m_\tau^2/\Lambda^2_{\MSsch})={1\over\beta_0}\left({1\over h}
-c_1\ln{1+c_1h\over h\beta_0}\right)+
\int_0^h d\xi\left({1\over\beta(\xi)} 
- {1\over\beta_{(2)}(\xi)}\right)
\]
where $\beta_{(2)}(h)=-\beta_0 h^2-\beta_1 h^3$.

It is seen from the results given in the Tables that the difference
between the exact value (summed up in all orders in $h$) and the naive
one becomes smaller with the improvement of the approximation i.e.
after taking into account the larger number of
corrections in $h$. It is explained by the obvious observation that
the leading term of the difference between the exact result and
the naive one has formally the next order in $h$ in comparison with
the kept terms (cf. the situation with the usual
renormalization group summation of the perturbative corrections).
Since the quantity $h$ (more precisely $r_H$~(\ref{13})) is small 
enough the estimate of the analytic continuation effects by means 
of the expansion of exact formulas~(\ref{5},\ref{14}) into the series 
in $h$ is valid numerically. For the value $12r_H=1$, however, 
the situation is essentially changed even numerically as it can be
seen from the last lines of the tables although the estimates remain 
obviously unchanged parametrically. Even for the moderate numerical 
values of $r_H$, for example, $12r_H=0.5$, however, the change of the
parameters $h$ and $\Lambda_{\MSsch}$ under taking into
consideration the next approximation in $h$ is comparable in the
magnitude with the changes influenced by the summation of the
analytic continuation effects. The accuracy of experimental
determination of the quantity $r_H$ is permanently getting
improved and after accomplishing the construction of and putting into
operation the $c-\tau$-factories~\cite{17} will be considerably
better. So, 
if $r_H=\bar{r}_H+\Delta r_H$ then at $\bar{r}_H=0.60$ and $\Delta
r_H=0.03$ one can reliably tell apart the exact and naive results for
$h$ and $\Lambda_{\MSsch}$.

In conclusion I would like to stress once again the main result of the
paper. The knowledge of the analytic and renormalization group
properties of the function $d(Q^2)$ allows to sum up all the regular
contributions into the moments of the spectral density $r(s)$ in all
orders of perturbation theory in the strong coupling constant
$h=\als(m_\tau)/4\pi$. The summation of such contributions at the
calculation of the inclusive semileptonic width of the $\tau$-lepton
is necessary for precise comparison of the parameters
$\Lambda_{\MSsch}$ extracted from the different processes
depending on different energy scales.

\newpage

\noindent
Table 1. The leading approximation 
$a_1=0$, $a_2=0$, $\beta_1=0$, $\beta_2=0$.
\vskip 0.3cm 

$
\begin{array}{lllll}
\hline \\
12 r_{H}^{\rm exp} & 10^2h^{(1)}&   10^2h^{(1)}_N&
\Lambda^{(1)}_{\MSsch}
&  \Lambda^{N(1)}_{\MSsch}\\ \\
\hline 
0.30 &1.97 &1.94 &107 &102 \\  
0.40 &2.51 &2.44 &196 &183 \\  
0.50 &3.04 &2.90 &287 &262 \\
0.60 &3.55 &3.32 &374 &334 \\
0.70 &4.08 &3.71 &457 &399 \\
1.00 &5.77 &4.76 &682 &555\\ 
\hline   
\end{array}
$

\vskip 1cm 

\noindent
Table 2. The next-to-leading approximation  $a_2=0$, $\beta_2=0$.
\vskip 0.3cm 

$
\begin{array}{lllll}
\hline \\
12 r_{H}^{\rm exp} & 10^2h^{(2)}&   10^2h^{(2)}_N&
\Lambda^{(2)}_{\MSsch}
&  \Lambda^{N(2)}_{\MSsch}\\ \\
\hline 
0.30 &1.74 &1.72 &161 &155 \\  
0.40 &2.17 &2.11 &277 &261 \\  
0.50 &2.57 &2.45 &390 &357 \\
0.60 &2.96 &2.75 &496 &441 \\
0.70 &3.36 &3.03 &596 &515 \\
1.00 &4.79 &3.74 &876 &692\\ 
\hline   
\end{array}
$

\vskip 1cm

\noindent
Table 3. The next-next-to-leading approximation.
\vskip 0.3cm 

$
\begin{array}{lllll}
\hline \\
12 r_{H}^{\rm exp} & 10^2h^{(3)}&   10^2h^{(3)}_N&
\Lambda^{(3)}_{\MSsch}
&  \Lambda^{N(3)}_{\MSsch}\\ \\
\hline 
0.30 &1.70 &1.70 &140 &140 \\  
0.40 &2.09 &2.07 &237 &232 \\  
0.50 &2.45 &2.39 &330 &316 \\
0.60 &2.81 &2.68 &416 &387 \\
0.70 &3.17 &2.95 &498 &448 \\
1.00 &4.69 &3.62 &751 &585\\ 
\hline   
\end{array}
$


\begin{thebibliography}{99}
\bibitem{1}  See, for example, J.D.~Bjorken, Preprint SLAC-PUB-5361,
T/E, 1990.
\bibitem{2} See, for example, G.~Altarelli, Preprint CERN-TH.5834/90, 1990;\\
C.~Jarlskog, Preprint CERN-TH.5740/90, 1990.
\bibitem{3} E.~Braaten, Phys. Rev. Lett. 60 (1988) 1606, 
Phys. Rev. D39 (1989) 1458.
%%CITATION = PHRVA,D39,1458;%%
\bibitem{4}  S.~Narison and A.~Pich, Phys. Lett. B211 (1988) 183.
%%CITATION = PHLTA,B211,183;%%
\bibitem{5}  A.~Pich, Preprint CERN-TH.5940/90.
\bibitem{6}  S.L.~Adler, Phys. Rev. D10 (1974) 132.
%%CITATION = PHRVA,D10,132;%%
\bibitem{7} R.~Shankar, Phys. Rev. D15 (1977) 755.
%%CITATION = PHRVA,D15,755;%%
\bibitem{8}  E.~Poggio, H.~Quinn and S.~Weinberg,
Phys. Rev. D13 (1976) 1958.
%%CITATION = PHRVA,D13,1958;%%
\bibitem{9} N.V.~Krasnikov, K.G.~Chetyrkin and A.N.~Tavkhelidze,
Phys. Lett. B76 (1978) 76.
%%CITATION = PHLTA,B76,76;%%
\bibitem{10} N.V.~Krasnikov, A.A.~Pivovarov and N.~Tavkhelidze,
Z. Phys. C19 (1983) 301.
%%CITATION = ZEPYA,C19,301;%%
\bibitem{11} A.A.~Pivovarov, Preprint INR P-707, 1991.
\bibitem{12} E.~Stueckelberg, A.~Peterman, Helv. Phys. Acta 26 (1953) 499;\\
M.~Gell-Mann, F.E.~Low, Phys. Rev. 95 (1954) 1300;\\
%%CITATION = PHRVA,95,1300;%%
N.N.~Bogoliubov, D.V.~Shirkov, Doklad. Akad. Nauk. SSSR
103 (1955) 391,\\ Nuov. Cim. 3 (1956) 845;
A.A.~Logunov, Zh. Eksp. Teor. Fiz. 30 (1956) 793;\\
L.V.~Ovsiannikov, Doklad. Akad. Nauk. SSSR 109 (1956) 1121.
\bibitem{13} O.V.~Tarasov, A.A.~Vladimirov and A.Yu.~Zharkov,
Phys. Lett. B93 (1980) 429.
%%CITATION = PHLTA,B93,429;%%
\bibitem{14} G.~Grunberg, Phys. Lett. B95 (1980) 70;\\
%%CITATION = PHLTA,B95,70;%%
A.L.~Kataev, N.V.~Krasnikov and A.A.~Pivovarov, 
Phys. Lett. B107 (1981) 115.
%%CITATION = PHLTA,B107,115;%%
\bibitem{15} K.G.~Chetyrkin, A.L.~Kataev and F.V.~Tkachov,
Phys. Lett. 85B (1979) 277;\\
%%CITATION = PHLTA,B85,277;%%
M.~Dine, J.~Sapirstein, Phys. Rev. Lett. 43 (1979) 668;\\
W.~Gelmaster, R.J.~Gonsalves, Phys. Rev. D21 (1980) 3112.
%%CITATION = PHRVA,D21,3112;%%
\bibitem{16} L.R.~Surguladze and M.A.~Samuel, OSU preprint, 1990;\\
S.G.~Gorishny, A.L.~Kataev and S.A.~Larin, ZhETP Lett. 53 (1991) 121.
%%CITATION = JTPLA,53,121;%%
\bibitem{17} J.~Kirkby, Preprint CERN-PPE/91-13, 1991.
\end{thebibliography}
\end{document}